\documentclass[epj]{webofc}
\usepackage[varg]{txfonts}  
\usepackage{xcolor}
\woctitle{MESON2018 - the 15$^\textrm{th}$ International Workshop on Meson Physics}
\begin{document}
\selectlanguage{english}
\title{Theoretical analysis of the $\gamma\gamma^{(*)} \to \pi^0\eta$ process}

\author{Oleksandra~Deineka\inst{1}\fnsep\thanks{\email{deineka@uni-mainz.de}} \and
        Igor~Danilkin\inst{1} \and
        Marc~Vanderhaeghen\inst{1} 
}

\institute{Institut f\"ur Kernphysik \& PRISMA  Cluster of Excellence, Johannes Gutenberg Universit\"at,  D-55099 Mainz, Germany
          }

\abstract{

The theoretical analysis of the $\gamma\gamma \to \pi^0\eta$ process is presented within the energy range up to 1.4 GeV. The $S$-wave resonance $a_0(980)$ is described involving the coupled channel dispersive framework and the $D$-wave $a_2(1320)$ is approximated as a Breit-Wigner resonance. For the $a_0(980)$ the pole is found on the IV Riemann sheet resulting in a two-photon decay width of $\Gamma_{a_0\to\gamma\gamma}=0.27(4)$ keV. The first dispersive prediction is provided for the single-virtual $\gamma\gamma^*(Q^2)\to\pi^0\eta$ process in the spacelike region up to $Q^2=1$ GeV$^2$.   
}
\maketitle
\section{Introduction}
\label{sec:intro}
The comprehensive study of the reaction $\gamma\gamma^* \to \pi^0\eta$ with one off-shell photon serves as important constraint to the hadronic light-by-light contribution to the anomalous magnetic moment of the muon $a_{\mu}$ \cite{Benayoun:2014tra}. Moreover, measuring photon-fusion processes with a single tagged technique is a part of two-photon physics program of the BESIII Collaboration \cite{Redmer:2018gah}. Currently, there is only real photon 
high-statistic data from Belle Collaboration \cite{Uehara:2009cf}, which provides the valuable information on the nature of two resonances: scalar $a_0(980)$ and tensor $a_2(1320)$. In particular, it allows to extract the two-photon strength of the $0^{++}$ isovector channel. 

The vast majority of the present models \cite{Oller:1997yg, Achasov:2010kk,Danilkin:2012ua} contain a significant number of parameters to be fitted to the experimental data and therefore, have a very limited predictive power. The proposed dispersive method \cite{Danilkin:2017lyn} complies with the fundamental properties of the $S$-matrix, i.e. analyticity and unitarity, and in principle allows a parameter-free description of both real and single-virtual processes.

\section{Formalism}
\label{sec:formalism}

We consider the process $\gamma(q_1,\lambda_1)\gamma^*(q_2,\lambda_2)\to\pi(p_1)\eta(p_2)$, in which one of the photons is real, while the second one has a spacelike virtuality $q_2^2=-Q^2$. 
For studying the low-lying  resonances, it is convenient to perform the partial wave (p.w.) expansion of the helicity amplitudes with the fixed isospin ($I$).

\subsection{Coupled-channel Omn\`es representation}
\label{sec:Omnes_representation}
Since both experimental and theoretical results serve in favor of $a_0(980)$ being a coupled-channel $\pi\eta$ and $K\bar{K}$ system, the inclusion of $K\bar{K}$ intermediate state appears to be necessary in order to describe the data. We follow the approach implemented in \cite{GarciaMartin:2010cw} for $\gamma\gamma \to \pi\pi$, where the well-known Born left-hand cut ($s<0$) is separated from the heavier $t$- and $u$-channel contributions ($s<s_L$), thus allowing to write a dispersion relation for the function $\Omega^{-1}(s)(h(s)-h^{\text{Born}}(s))$, which contains both left- and right-hand cuts, 

\begin{align}\label{disp}
\left(\begin{array}{c}h^{0}_{1,++}\\
k^{0}_{1,++}\end{array}\right)=\left(\begin{array}{c}0\\
k^{0, \text{ Born}}_{1,++}\end{array}\right) +\Omega_{1}^{0}(s) &\Bigg[\left(\begin{array}{c}a\\b\end{array}\right) +\frac{s-s_{th}}{\pi}\int\limits_{-\infty}^{s_L}
\frac{ds'}{s'-s_{th}}\,\frac{\Omega_{1}^{0}(s')^{-1}}{s'-s}\left(\begin{array}{c}\text{Disc}\,h^{0,\,\text{V exch}}_{1,++}(s')\\\text{Disc}\,k^{0,\,\text{V exch}}_{1,++}(s')\end{array}\right) \nonumber \\
 &-\frac{s-s_{th}}{\pi}\int\limits_{s_{th}}^{\infty}\frac{ds'}{s'-s_{th}}\,\frac{\text{Disc}\,\Omega_{1}^{0}(s')^{-1}}{s'-s}\left(\begin{array}{c}0\\ k^{0,\text{ Born}}_{1,++}(s')\end{array}\right)
\Bigg],
\end{align}
where $h^{0}_{1,++}\,(k^{0}_{1,++})$ denote the $I=1$, $S$-wave, helicity-0 amplitude of $\gamma\gamma^{(*)}\to \pi\eta \, (K\bar{K})$ process and $s_{th}=(m_{\pi}+m_{\eta})^2$. In the general case, both subtraction constants ($a, b$), $s_L$ and the p.w. amplitudes have a $Q^2$ dependence.
Due to the absence of experimental data on $\pi\eta$ and $K\bar{K}$ scattering, the data-driven dispersive determination of the Omn\`es matrix is impossible. Instead, we use the approach accurately described in \cite{Danilkin:2011fz}, which is based on the $N/D$ ansatz and contains only few known relevant parameters. \

\subsection{Left-hand cuts}
\label{sec:Left-hand_cuts}

While the Born terms can be calculated from the scalar QED, the other contributions can be approximated by vector-meson exchange diagrams. We use the simplest Lagrangian which couples photon to pseudoscalar (P) and vector (V) meson fields: 
\begin{equation}\label{lagrangian}
\mathcal{L}_{VP\gamma}=e C_V \epsilon^{\mu\nu\alpha\beta}F_{\mu\nu}\partial_{\alpha}V_{\beta} \,,
\end{equation}
where $C_V$ are the radiative couplings fixed to the experimental values \cite{Patrignani:2016xqp} of partial width of light vector mesons. Naturally, $t$- and $u$-channel vector meson exchange terms lead to the logarithmic behavior of the p.w. amplitudes, from which the position of the closest left-hand cut $s_L$ can be extracted.
For the single-virtual process the left-hand cuts structure is more complicated. Unlike in the real case, $t$- and $u$-channels left-hands cuts are different. Moreover, the additional left-hand cuts intervals appear in each channel \cite{Moussallam:2013una}, i.e.
$[s_{L,t}^{+},0]$ and $[0,s_{L,u}^{+}]$ which have to be taken into account.  

In contrast to the full p.w. amplitudes, their imaginary parts, which enter the dispersion relation Eq.~(\ref{disp}) along the left-hand cuts are asymptotically bounded  $\text{Im }h_{++}^{0,\,\text{V exch}}(s\to-\infty)\to\text{const}$. Since the considered Omn\`es function is bounded at high energies, one subtraction is sufficient for the convergence. The appearing subtraction constants can be fixed by matching the Lagrangian-based amplitudes to the dispersive result at the threshold, thus leading to the completely parameter-free description.

\subsection{Form factors}
\label{sec:form_factors}

Regarding the single virtual process, we have to consider the transition form factors (TFFs). The kaon and pion TFFs are described by the monopole fit to the available data. However, there is no experimental data for the vector mesons form factors of a type $F_{VP\gamma}(Q^2)$ in the spacelike region and therefore, we favor the simple VMD \cite{Sakurai} prediction. We note also, that the difference between TFFs of $\{\rho\pi\gamma, \rho\eta\gamma, \omega\pi\gamma, \omega\eta\gamma\}$ is negligible in the region up to $Q^2=1 \text{ GeV}^{2}$. In order to allow the conservative variation, we take $F_{\rho\pi\gamma}(Q^2)$ as the central value and include the deviation up to the $\{K^{*,\pm}K^{\pm}\gamma, K^{*,0}K^0\gamma\}$ TFFs as the error source. The exisiting dispersive analyses of the $\omega\pi\gamma$ TFF \cite{Schneider:2012ez,Danilkin:2014cra} lie within this range.
The TFF of $a_2(1320)$ is given by two estimates:
it either can be taken equal to the pion TFF  using light-by-light sum rule arguments \cite{Pascalutsa:2012pr}, or to the form factor of the isoscalar counterpart $f_2(1270)$, which has been recently measured by Belle \cite{Masuda:2015yoh}. 
We take the spread between these two results as our error estimate.

\section{Results}
\label{sec:results}

\subsection{Pole position and two photon coupling of $a_0(980)$}
\label{sec:pole}

In order to identify the properties of $a_0(980)$ resonance, we performed the analytical continuation of the hadronic amplitudes to the complex $s$-plane and found a pole on the IV Riemann sheet (Im$p_{\pi\eta}>0$, Im$p_{K\bar{K}}<0$) \cite{Danilkin:2017lyn}: \small{$\sqrt{s^{\text{IV}}_{a_0}}=\left(1.12^{-0.07}_{+0.02}\right) -\frac{i}{2}\left(0.28^{+0.08}_{-0.13}\right)$} \normalsize GeV, where the errors reflect the variation of the conformal mapping paramter $\Lambda_S$, which sets the scale from where on the s-channel physics is integrated out. In this case, the $a_0(980)$ appears as a sharp peak exactly at the two kaon threshold. The residue of the pole gives the couplings ratio $|c_{K\bar{K}}/c_{\pi\eta}|=0.98^{-0.07}_{+0.20}$ which indicates a strong coupling to both $\pi\eta$ and $K\bar{K}$ channels, as expected. In the narrow-width approximation, the radiative width of $a_0(980)$ is determined as
\begin{equation}\label{Gamma_a0}
\Gamma_{a_0\to\gamma\gamma}=\frac{|c_{\gamma\gamma}|^2}{16\pi M_{a_0}}=0.27(4)\,\text{keV}\,.
\end{equation}

\begin{figure}[t]
\centering
\includegraphics[scale=0.52]{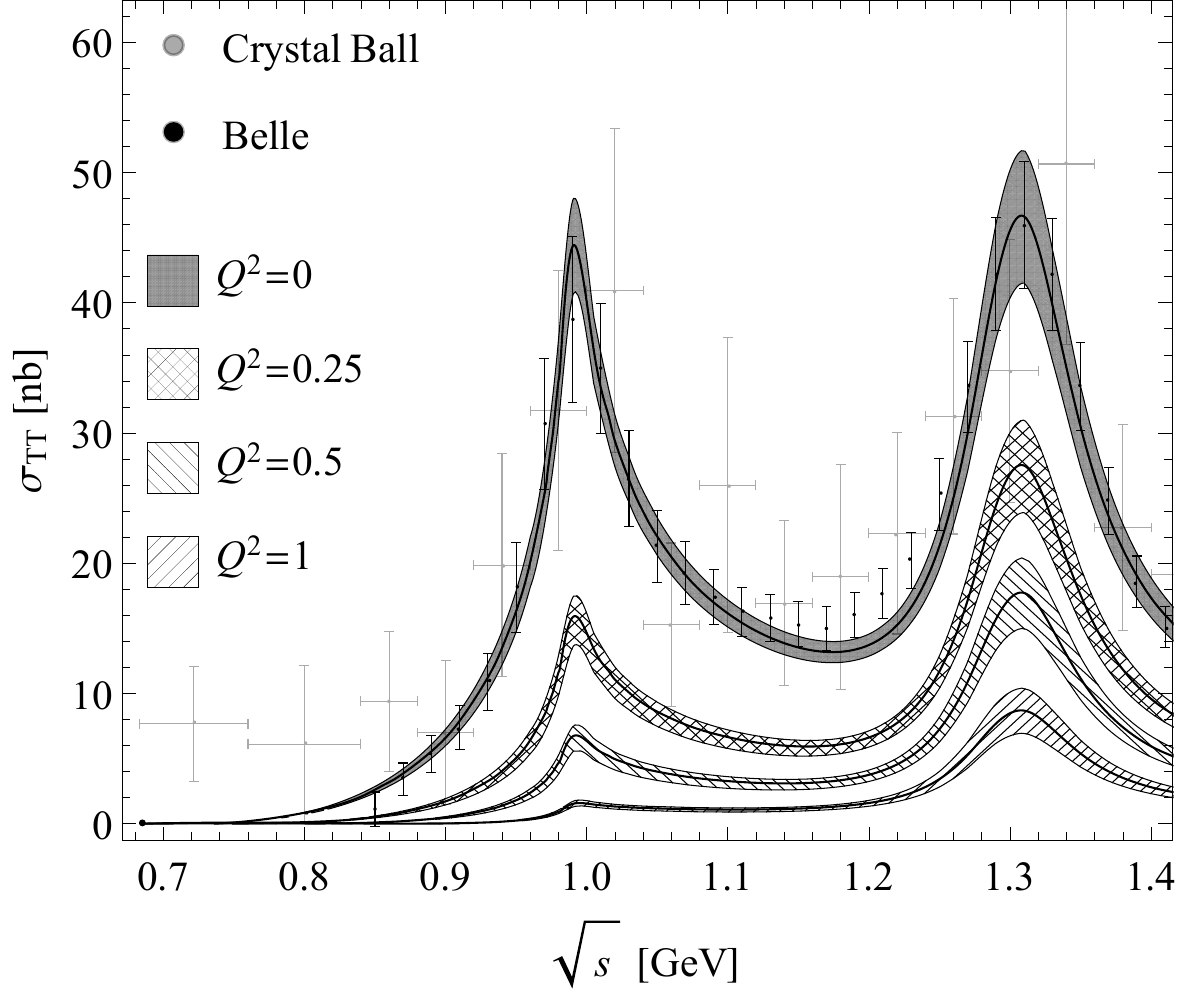}
\hfill
\includegraphics[scale=0.54]{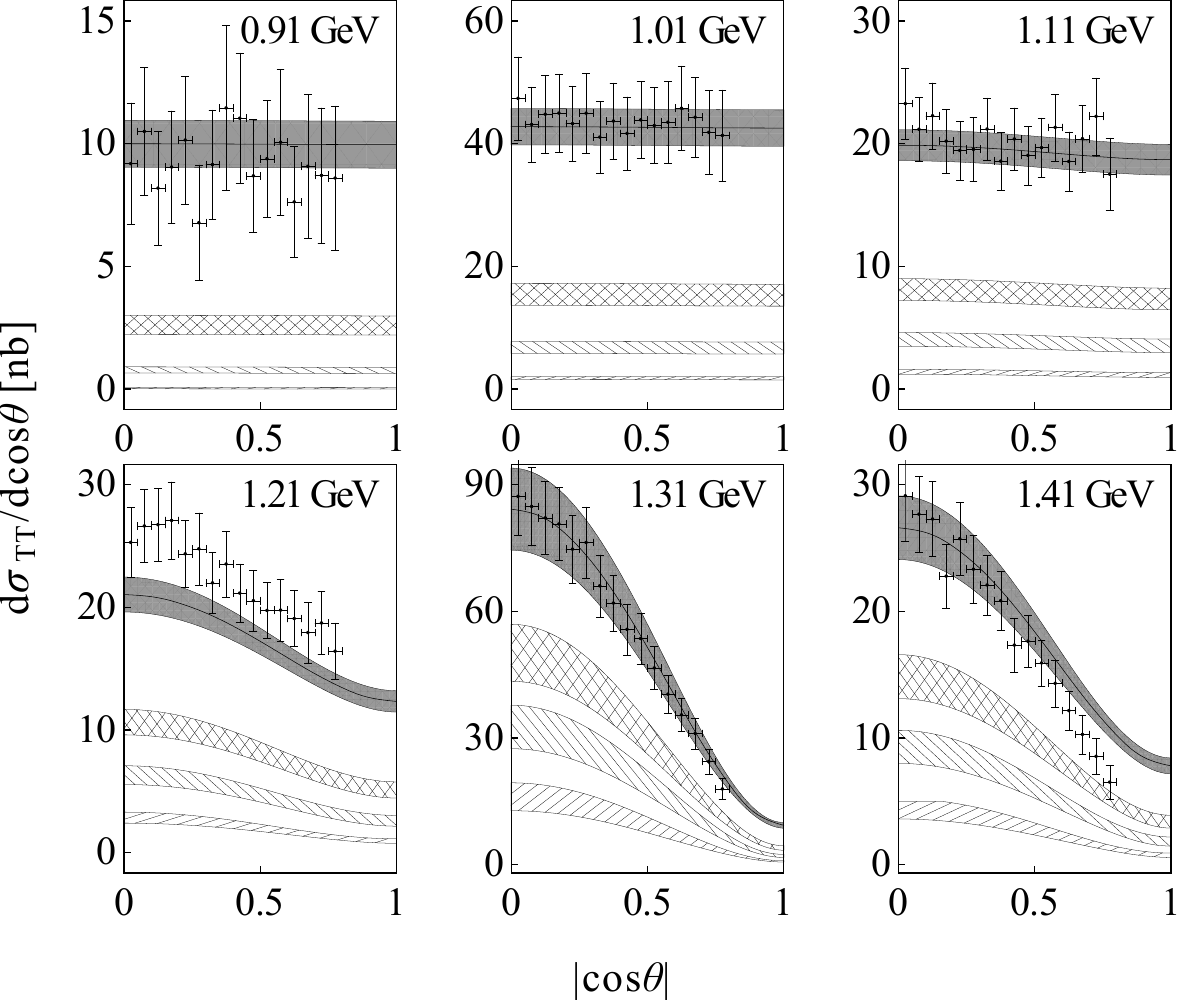}
\caption{Total ($|\cos\theta|<0.8$) and differential cross sections for the $\gamma\gamma^*\to\pi^0\eta$ process. The error bands reflect the uncertainties of the universal coupling $VP\gamma$, the variation of $\Lambda_S$, couplings of the $a_2(1320)$ to $\gamma\gamma, \pi\eta$ and the variation of the TFFs as is described in Sec.~\ref{sec:form_factors}. The data are taken from \cite{Uehara:2009cf, Antreasyan:1985wx} }
\label{fig:cross_section}     
\end{figure}

\subsection{Cross sections of the $\gamma\gamma^{(*)}\to\pi\eta$ processes}
For the analyses of the total and angular cross sections, we take the $a_2(1320)$ resonance explicitly within the assumption that it is predominantly produced by the helicity-2 state (similar to $f_2(1270)$ in \cite{Drechsel:1999rf}). Together with the proposed dispersive method for $a_0(980)$ this yields a parameter-free description of the $\gamma\gamma\to\pi^0\eta$ cross section \cite{Danilkin:2017lyn}, which is in the reasonable agreement with the data from the Belle Collaboration \cite{Uehara:2009cf}. However, the comparatively large uncertainties can be further narrowed down. We used the two-photon invariant mass distribution of the cross channel $\eta\to\pi^0\gamma\gamma$ decay data \cite{Nefkens:2014zlt} in order to further constrain the left-hand cuts. Then, making use of the high-statistic data on the total cross section \cite{Uehara:2009cf} we perform the fit in the region $\sqrt{s}<1.1 \text{ GeV}$  to narrow down the uncertainty of the conformal mapping parameter $\Lambda_S$, entering in the description of the hadronic final state interaction.

For the single virtual process we restrict ourselves to the values of $Q^2=0.25, 0.5, 1$ GeV$^2$ and the case of $\sigma_{TT}$, i.e. when both photons are transversal. In contrast to the real case, there is also a non-zero $P$-wave amplitude $h^1_{1,++}(s)$, which gives a negligible contribution to the cross section compared to the combined result of the $S$- and $D$-waves shown on Fig.~\ref{fig:cross_section}. The resulting cross section $\sigma_{TT}$ should be further confronted with the future data from BESIII \cite{Redmer:2018gah}, as well as the $\sigma_{TL}$, where one photon is longitudinally polarized. The latter is the part of an ongoing analysis.

\end{document}